# *Electronic Defects in Metal Oxide Photocatalysts*

*Ernest Pastor[1*], Michael Sachs[2], Shababa Selim[2], James R. Durrant[2], Artem A. Bakulin[2] and Aron Walsh[3,4*]*

[1] ICFO-Institut de Ciències Fotòniques, The Barcelona Institute of Science and Technology, Av. Carl Friedrich Gauss 3, 08860 Castelldefels (Barcelona), Spain

[2] Department of Chemistry and Centre for Processable Electronics, Imperial College London, White City Campus, London W12 0BZ, UK

[3] Department of Materials, Imperial College London, London SW7 2AZ, UK

[4] Department of Material Science and Engineering, Yonsei University, Seoul 03722, Korea

*ernest.pastor@icfo.eu, a.walsh@imperial.ac.uk

**Abstract |** A deep understanding of defects is essential for the optimisation of materials for solar energy conversion. This is particularly true for metal oxide photo(electro)catalysts, which typically feature high concentrations of charged point defects that are electronically active. In photovoltaic materials, except for selected dopants, defects are considered detrimental and should be eliminated to minimise charge recombination. However, photocatalysis is a more complex process where defects can play an active role, for example, by stabilising charge separation and mediating rate-limiting catalytic steps. Here, we review the behaviour of electronic defects in metal oxides, paying special attention to the principles underpinning the formation and function of trapped charges in the form of polarons. We focus on how defects alter the electronic structure, statically or transiently upon illumination, and discuss the implications of such changes in light-driven catalytic reactions. Finally, we consider the applicability of lessons learned from oxide defect chemistry to new photocatalysts based on carbon nitrides, polymers and metal halide perovskites.

# Introduction

Solid-state photo(electro)chemistry uses semiconductors to capture sunlight and drive useful chemical reactions, such as the production of $H_2$ from water or the reduction of $CO_2$ into higher energy density compounds like ethanol. Great progress has been made since the pioneering work by Gerischer studying semiconductor-liquid interfaces[1] and the milestone report of water splitting in $TiO_2$ by Fujishima and Honda,[2] making today's semiconductor photo(electro)chemistry an increasingly viable technology. The advances have been such that current research is no longer only focused on demonstrating catalysis but is working to push conversion efficiencies to an economically competitive range and develop working devices.[3–6]

The biggest strides in the field have been driven by improvements of well-known metal oxides, most notably through nanostructuring (e.g "cauliflower" $\alpha$-$Fe_2O_3$ reported by Grätzel and co-workers[7]) , the assembly of heterojunctions (e.g. $WO_3$:$BiVO_4$)[8] and the discovery of more efficient compositions (e.g. $SrTiO_3$:Al nanoparticles reported by Domen and co-workers[9]). Critically, all these advances have been underpinned by improvements in the control of the concentration and distribution of defects, which even in small amounts can radically change the properties of a solid and the performance of a device. Defects are responsible for a variety of phenomena detrimental to the electronic properties of energy-conversion materials. For example, the structural deformations induced by defects can hinder charge transport and the electronic deformations can act as recombination centres compromising device performance[10] (e.g. quenched luminescence in Co-doped ZnO). However, despite their unappealing name, defects can also have a desirable influence on the material properties. For example, chemical doping can turn an insulating oxide into an efficient photocatalyst by increasing carrier concentrations (e.g. enhanced electron concentrations in Nb-doped $TiO_2$),[11] concomitantly generating ionised dopants as point defects in the structure. Similarly, control of electronic states associated with defects can help tune and extend charge carrier lifetimes.[12,13] Over recent years, there has also been increasing awareness of the importance of polarons in the operation of photo(electro)catalytic devices, namely charges which trap through lattice deformations even in the absence of a physical point defect.[14,15]

Understanding the role of defects and localised carriers that form polarons is helpful for improving functional solids for a broad range of applications, but is absolutely critical in the case of heterogeneous photocatalysts. In such systems light triggers the reorganisation of chemical bonds at a solid-liquid (or sometimes solid-gas) interface, requiring the participation of surface dangling bonds (i.e. surface defects) for operation. Even more importantly, defect formation is in itself a chemical reaction which depends strongly on the chemical nature of the solid. Complications emerge when the materials properties required to favour a desired chemical transformation are not so different from those that favour defect formation.

Here we review the impact that electronic defects have on the function of photo(electro)catalytic oxides. We use the term photocatalysis to refer to both photocatalytic and photo(electro)catalytic processes. While the operation principles are different in both cases (see Box 1), here we emphasise the common role that defects play in such technologies. By electronic defects, we refer to changes in the electronic structure induced by point defects (e.g. oxygen vacancies) and/or polaronic states associated with charge carriers. While oxide photocatalysis is an extremely rich field where promising materials are rapidly being discovered[4,16–26] and new reactions being catalysed (e.g. $CO_2$ reaction, pollutant degradation)[27–30] both in liquid and gas phase,[31–33] here we focus primarily on established cases of photocatalytic oxides employed for water splitting. In particular, we use $Fe_2O_3$,[34–37] $TiO_2$,[38–40] $WO_3$[41–43] and $BiVO_4$[5,44–46] as examples to provide a framework to understand how defect control can help improve photocatalytic yields in oxide-based systems.

The manuscript is divided as follows: first we review the defect chemistry in metal oxides. We pay special attention to the formation of structural imperfections in the form of vacancies and polarons, and show how such processes alter the fundamental electronic structure of the catalysts. Secondly, we outline the key processes that light-driven catalysts must overcome from the initial equilibration with the liquid phase to the product formation. Next, we discuss how chemical imperfections impact such processes, making emphasis on how sub-band gap energy levels affect the optoelectronic properties of the metal oxide and the kinetics and energetics of the reactive excited state. Finally, we consider the applicability of the lessons learned from the defect chemistry in oxides to emerging photocatalyst based on metal halide perovskites, as well as carbon nitride and conjugated polymers.

## **Point defects in metal oxides**

The formation of defects in solids is unavoidable and results in the loss of the translational symmetry of the crystallographic unit cell. Symmetry breaking can happen across multiple dimensions, giving rise to 3D volumetric defects (e.g. pores), 2D planar defects (e.g. grain boundaries), 1D linear defects (dislocations), and 0D point defects (e.g. vacancies).[47,48] While all types of defects can affect the electronic and catalytic properties of metal oxides,[49–51] herein we focus on point defects as they are the most intrinsic in nature and play a major role in the optimisation of catalytic performance. For a comprehensive overview of the classification of point defects in crystals and associated characterisation techniques, we refer to the seminal work of Kröger.[52] We first provide an overview of the electronic structure of metal oxides and use it as a framework to discuss the chemical principles that underpin the formation of imperfections (both in the dark and upon illumination).

**Electronic structure and band gaps.** The electronic structure and the magnitude of the band gap of metal oxides is influenced by the crystal symmetry, the ionicity of the metal-oxygen bond, and the degree of orbital mixing or hybridisation.[53,54] In general, ionic oxides with a large energy difference between metal and oxygen atomic orbitals tend to exhibit large band gaps outside the range for solar energy conversion.[55] This is the case of ZnO (3.4 eV), Ga$_2$O$_3$ (4.8 eV) and Al$_2$O$_3$ (8.8 eV) with an electronic band structure formed upon mixing of oxygen *2p* and metal *s* orbitals. In contrast, transition metal oxides (TMOs) with partially filled valence *d* orbitals, such as Fe$_2$O$_3$ (2.2 eV) sustain additional bonding interactions involving the metal *d* and/or *s* orbitals, which lower the band gap. Often such changes are sufficient to absorb photons in the visible part of the solar spectrum, making these metal oxides prime candidates for photocatalysis.

As a first approximation, the electronic structure of most TMOs can be explained from a crystal-field perspective in which the oxygen anions are arranged in different geometries around a metal cation. Electrons that occupy the *d* orbitals of the metal centre experience different degrees of repulsion depending on the orbital orientation with respect to the surrounding oxygen anions. In an octahedral geometry ($O_h$), metal atomic orbitals (AO) with $t_{2g}$ symmetry ($d_{xy}, d_{yz}, d_{xz}$) are stabilised, while AO with $e_g$ symmetry ($d_{x^2-y^2}, d_{z^2-r^2}$) are destabilised (**Figure 1a**). The removal of an oxygen atom from the octahedron, for example due to an oxygen vacancy or dangling bond at the surface, leads to a square pyramidal symmetry ($C_{4v}$) with further splitting and stabilisation of the $t_{2g}$ and $e_g$ orbitals as shown in **Figure 1a**.[56,57]

While crystal field theory provides an initial estimate of the electronic structure, this model does not describe covalent interactions. This extension is achieved with ligand field theory that considers the overlap and mixing of oxygen *2p* and metal *3d* wavefunctions. Within this molecular orbital perspective, in an octahedral ligand field, metal atomic orbitals with $t_{2g}$ symmetry have weak overlap with ligand oxygen O *2p* orbitals forming π-bonding and π*-antibonding molecular orbitals (MO). In contrast, metal AO with $e_g$ symmetry strongly interact with oxygen orbitals forming σ-bonding and σ*-antibonding MO with mixed metal/oxygen character. The remaining O *2p* orbitals that do not interact with *d* states form σ$_o$- non-bonding orbitals with the same energy as the original AO. **Figure 1b** shows a molecular orbital diagram for the building blocks of a transition metal oxide. The nature of the frontier orbitals will determine the redox chemistry and tendency for defect formation.

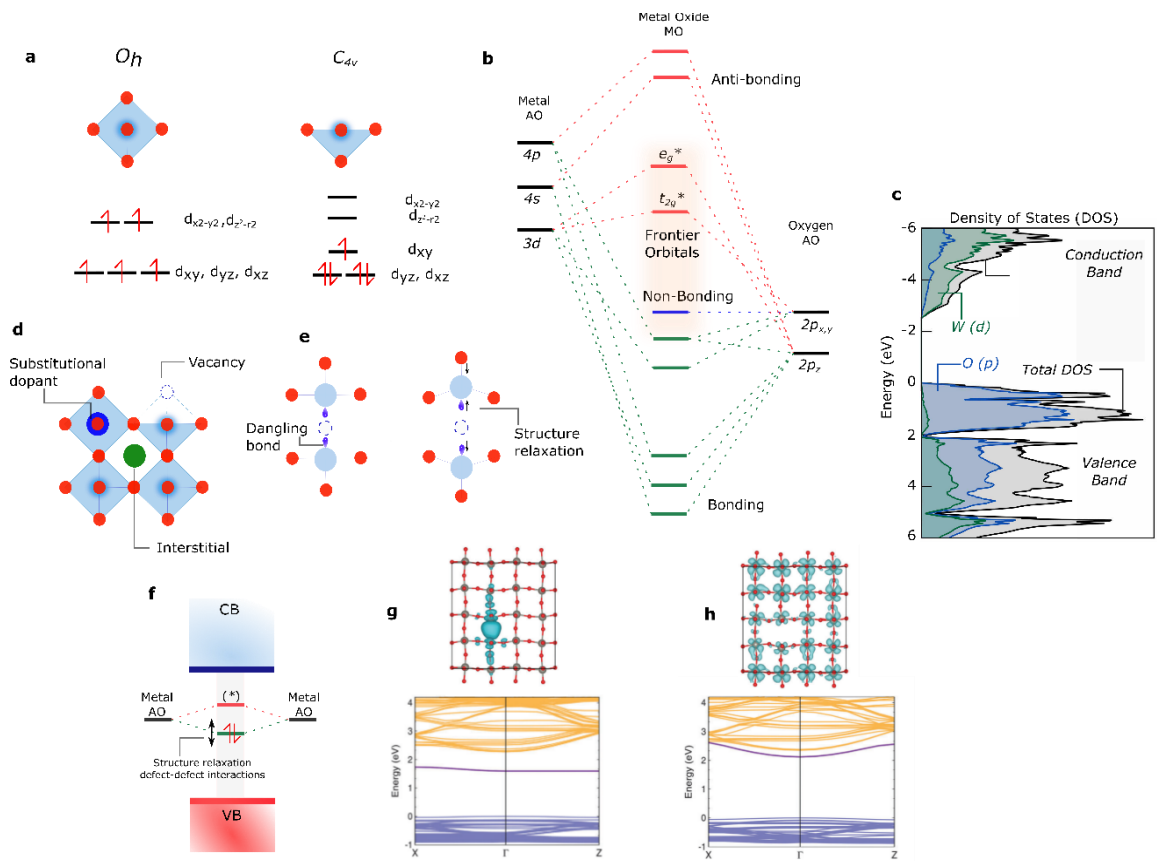

**Figure 1: Point defects in a metal oxide. (a)** Crystal field splitting of *d*-electrons in an octahedral ($O_h$) and square pyramidal ($C_{4v}$) coordination environment. **(b)** Molecular orbital diagram describing the building blocks of a transition metal oxide. **(c)** Electronic density of states calculated for $WO_3$. **(d)** Some types of point defects that form in metal oxides. **(e)** Oxygen vacancy in an oxide where the loss of a dangling bond upon ionisation results in a structural relaxation. **(f)** Electronic band diagram of an oxygen vacancy in a metal oxide. The defect level is close in energy to the metal atomic orbitals and can be shifted by structural relaxations, as well as defect-defect interactions, **(g)** charge density of a deep vacancy state and corresponding band structure and **(h)** charge density of a shallow vacancy state and corresponding band structure (adapted from[58]). Relaxation of the local structure around the vacancy can induce a transition from a localised to delocalised (deep to shallow) state.

**Frontier orbitals.** While it can be straightforward to define a small set of orbitals as the frontier orbitals (HOMO and LUMO) in a molecule which are responsible for the chemical reactivity, in a macroscopic crystal the picture is complicated by crystal momentum and the associated band dispersion in *k*-space. However, we can retain a similar picture by considering groups of bands. Electronic bands are generally grouped as a function of energy and depicted as a density of states (DOS) as shown in **Figure 1c** for $WO_3$. The states at the edges or maxima (minima) of the valence and conduction bands (VBM and CBM, respectively) correspond to the frontier crystal orbitals of the metal oxide. The bonding/antibonding nature, as well as the spatial distribution of those states will dictate the behaviour of the oxide when electrons are added (removed) or when dangling bonds are formed either in the bulk or at the surface. Hence, similar to molecular systems, the nature of the VBM and CBM will determine chemical reactivity but also influence the response to defect formation. Indeed, such analysis has been successful in describing physical phenomena in oxides, as well as reactivity trends.[59,60]

As a general rule, increasing the metal cation electronegativity (i.e. higher oxidation state and fewer $d$ electrons) brings the valence $d$ states energetically closer to the O $2p$ states and enhances the oxygen-metal hybridisation. If the metal valence orbitals are similar in energy to O $2p$, then the upper valence band will be a filled antibonding state, as found for the $d^{10}$ case of $Cu_2O$.[61] For transition metal oxides employed in photocatalysis such as $WO_3$, $BiVO_3$ or $Fe_2O_3$ the CBM is primarily composed of antibonding metal-like $d$-orbitals while the VBM typically consists of a hybridisation of metal $d$ states with oxygen $2p$ orbitals of bonding character or non-bonding states.[62] Such energetic make up has important consequences in the tendency to form point defects, localise charge in the form of polarons and ultimately in photocatalytic yields. For example, the $d^0$ conduction band of $TiO_2$ will become occupied by excess electrons, which can form localised or delocalised charges depending on the crystal polymorph.[63]

**Types of point defects.** Points defects can be broadly classed in two types: (1) *intrinsic* species such as vacancies, where an atom is missing from a regular crystal site, or interstitials, where an atom occupies a crystal site that is normally empty (**Figure 1d**); (2) *extrinsic* species, including unintentional incorporation of impurities or deliberate doping.[64] As shown below, trapped electrons in the form of polarons can also be considered a distinct type of electronic defect.

**Concentration of point defects.** A thermodynamic analysis can yield insights into the concentrations of particular defect species that we may expect in a crystal. The configurational entropy (ΔS) of a static perfect crystal is zero, as each atom is placed in its ideal crystallographic site. The consequence is that at non-zero temperatures the free energy of a crystal (ΔG=ΔH-TΔS) can be decreased by introducing configurational entropy in the form of defects.[47] A balance emerges between the enthalpic cost of breaking chemical bonds and the entropic gain of forming more imperfections. The resulting exponential dependence means that the equilibrium concentration (c) of a defect species is highly sensitive to its energy of formation (ΔH$_f$)[64,65]:

$$c = N \exp\left(-\frac{\Delta H_f}{k_B T}\right)$$

(*eq 1*)

where $N$ represents the density of associated atomic sites in the crystal.

What makes the situation more complex is that the defect formation energy is not a constant, but is a function of the atomic and electronic chemical potentials. In this way, the concentrations of different defect species are coupled and the defect profile of a particular sample will depend on its history, including the growth environment (e.g. oxygen partial pressure) and thermal processing (e.g. annealing time or quenching rate).[66–68] Let us take the example of the oxygen vacancy, a ubiquitous defect species in metal oxides. Vacancies can be formed by exchange of oxygen with the atmosphere as described by $O_O \rightleftharpoons V_O + \frac{1}{2}O_2(g)$. Each oxygen vacancy has the potential to be a double donor ($V_O^{2+}$) that will transfer two electrons to the conduction band (n-type donor).[69–71] The formation energy and equilibrium concentration of this defect will be sensitive to the annealing temperature, the partial pressure of oxygen, as well as the presence of other electron donating defects in the host crystal. Raising the partial pressure during high temperature annealing and the presence of an electron rich environment (n-type) will increase the formation energy and thus reduce the concentration of oxygen vacancies in the bulk. The semiconductor $Fe_2O_3$ has a concentration of oxygen sites in the crystal of around $N = 6 \times 10^{22}$ cm$^{-3}$. Assuming a typical defect formation energy of $\Delta H_f = 1$ eV, an annealing temperature of $T = 873$ K will yield a moderate net vacancy concentration of $c = 6 \times 10^{17}$ cm$^{-3}$.

**Energy levels in the band gap.** From an electronic viewpoint, the biggest consequence of defect formation is a change in the electroneutrality conditions of the metal oxide. An ideal stoichiometric material has a filled valence band and an empty conduction band, which are separated by a band gap sufficiently large to supress thermal excitations of electron and hole charge carriers. The incorporation of charged defects is associated with the creation of new energy levels in the band gap. The position of sub-band gap defect levels is related to energy required to exchange charge carriers (electrons or holes) with the valence and conduction bands. For example, a singly-charged oxygen vacancy ($V_O^+$) may be further ionised to inject an electron into the conduction band:

$$V_O^+ \rightarrow V_O^{2+} + e^-$$

(*eq 2*)

or it may arrive in the same final charge state by capturing a hole from the valence band:

$$V_O^+ + h^+ \rightarrow V_O^{2+}$$

(*eq 3*)

The energy required for each ionisation (ΔE) corresponds to the separation between the defect level and the valence or conduction band edge of the crystal. The location of ionisation levels in a semiconductor band gap is of major interest for the optimisation of energy conversions systems as it determines the beneficial or detrimental role of the defect. We distinguish two cases: a *shallow state* describes the case where a defect level lies close (usually within $k_BT \sim 26$ meV at room temperature) to the band edge and can be thermally ionised. A shallow state is often associated with a defect wavefunction that is delocalised over 10s-1000s unit cells.[72] In the hydrogenic limit, for a charged defect in a dielectric host, the shallow state energy is determined by the effective mass and dielectric constant of the host (in atomic units):

$$\Delta E = \frac{m^*}{2\varepsilon^2}$$

(*eq 4*)

Values of $\Delta E$ = 1–50 meV are common in tetrahedral semiconductors such as Si and ZnO.[72] As a consequence, shallow states typically support doping by increasing the concentrations of free carriers in the bands and thus the electrical conductivity.

In contrast, a *deep state* refers to the case where the ionisation energy is larger than thermal energy and is often associated with spatially localised defect wavefunctions on one or a small group of atoms. Deep states are usually formed following a large chemical perturbation, *e.g.* a physical vacancy in the structure or an elemental substitution with a large size or orbital mismatch (e.g. replacing O by As in a metal oxide). They are often associated with large structural distortions. In optoelectronic and photocatalytic devices, deep states can act as centres for trapping or recombination of charge carriers and may also pin the Fermi level to a limited energy range inside the band gap as discussed below.

We can make a further distinction concerning the defect levels. Optical levels correspond to a rapid vertical excitation (i.e. subject to the Franck–Condon principle), where there is no change in the final defect geometry. On the other hand, *thermal* levels correspond to a slower process, where the defect geometry has time to transform, as found in capacitance techniques including deep-level transient spectroscopy.[73] As a consequence, optical levels are deeper than the corresponding thermal levels by an amount corresponding to the structural relaxation energy of the particular defect species.

From a chemical viewpoint, the depth in the band gap of the defect states can be understood in terms of the frontier orbitals of the host crystal and the coordination environment of the defect site. The formation of, for example, an oxygen vacancy will create a void with dangling bonds on the adjacent metals that, to a first approximation, will have an energy similar to the metal atomic orbitals (**Figure 1e** and **1f**). Consequently, the formation of shallow defect states will be favoured when the associated orbitals are located energetically close to the edge of the bands or even within the bands.[74] In addition, weak hybridisation between the dangling metals, favoured by a large metal-metal distance and a low coordination environment at the defect site, will favour shallow defect formation.[75]

**Figure 1f** exemplifies this behaviour for the case of WO$_3$. The generation of an oxygen vacancy results in two dangling W bonds with an additional electron each (i.e. reduction of W(VI) to W(V)). This configuration leads to the formation of a doubly occupied energy level deep in the gap. However, bond length relaxation around the vacancy to minimise Coulomb repulsion increases the W-W distance and pushes the defect level closer to the conduction band and facilitates the injection of carriers.

**Figure 1g** shows the case of where charge is localised around the vacancy and a deep level is formed. In the absence of a strong structural distortion, the charge delocalises across neighbouring sites concomitantly forming a shallow defect level (**Figure 1h**).[58] When the continuum of states and the wavefunction of the defect overlap, the defect is considered to be resonant with the band.[65]

**Polarons.** An important consequence of point defect formation is the introduction of excess charge (electrons or holes) into a crystal. For example, doping $Fe_2O_3$ with Ti or Sn,[76] and $BiVO_4$ with W,[77] increases the concentration of electrons in the host material and enhances the n-type character. While defects can induce structural relaxations to accommodate the introduction of the dopant atoms, the excess electronic charge can also induce its own lattice distortion.[78,79] In particular, if the solid displays a strong dielectric response (i.e. it is highly polarisable), the excess charge can lower its energy by displacing the surrounding atoms through electron-phonon interactions, forming a quasi-particle known as a *polaron*.[78–83] As with other point defects, a polaron will introduce an energy level in the band gap (**Figure 2a** and **2b**). If the charge and its structural distortion are strongly bound to another defect (e.g. vacancy), the polaron can also modify this defect level.[55] For example, electron polarons in $BiVO_4$ localised in V centres change the effective oxidation state from V(V) to V(IV) and increase the V-O length by ~0.1Å.[84] Attaining such distortion might require an activation energy (**Figure 2b).** In addition, different configurations might exist for polarons near a defect (vacancy) as shown in **Figure 2c** for $BiVO_4$,[85] complicating the characterisation of the localised state both at the bulk and the surface.

From an energetic viewpoint, the driving force behind polaron formation is a balance between the potential energy gain upon distorting the structure and the kinetic energy loss associated with carrier localisation.[86] Minimisation of the total energy of the additional charge in a dielectric continuum yields the polaron binding energy as[86]:

$$E_P = \frac{1}{4\pi^2} \frac{m^* e^4}{2\hbar^2 \varepsilon_{eff}^2}$$

*(eq 5)*

Where m* is the effective mass, e and the electronic charge and $\varepsilon_{eff}$ is obtained from the static and ionic dielectric constants as:

$$\frac{1}{\varepsilon_{eff}} = \frac{1}{\varepsilon_\infty} - \frac{1}{\varepsilon_o}$$

*(eq 6)*

The polaron binding energy refers to the energetic stabilisation of a polaronic state with respect to a bare valence band hole or conduction band electron. Microscopically polaron formation is dictated by the coupling between charge carriers and the vibrational degrees of freedom. Within the Landau-Pekar model,[87] which considers only the characteristic frequency of a longitudinal optical phonon ($\omega_o$), the strength of this coupling, $\alpha_F$, is:

$$\alpha_F = e^2 \left(\frac{1}{\varepsilon_\infty} - \frac{1}{\varepsilon_o}\right) \sqrt{\frac{m^*}{2\omega_o}}$$

*(eq 7)*

Depending on the strength of the electron-phonon coupling, and the extent of the structural distortion, polarons are categorized as:[80] (i) *large polarons* (or Fröhlich polarons) which exhibit weak coupling and the distortion extends over more than a single unit cell and, (ii) s*mall polarons* (Holstein polarons) characterised by strong coupling and distortions that expand over less than a single unit cell with wavefunctions localised to dimensions on the order of interatomic distances.

Polaron formation is favoured in systems with large dielectric constants and heavy effective carrier masses (Equation 5 and 7). Evaluation of the electronic structure can therefore provide an indication of the tendency to form polarons. In general, poor overlap between atomic orbitals will result in weakly dispersive bands and large effective masses for carriers that will favour localisation.[55] Often metal oxides fall into this regime with frontier orbitals that are made up of weakly interacting orbitals and have high density of states around the band edges.

Such electronic make-up favouring polaron formation often occurs in transition metal oxides.[55] Indeed one of the most studied systems in photocatalysis is $TiO_2$ composed of Ti(IV) *$3d^0$* ions and in which electron paramagnetic resonance (EPR) studies have demonstrated the existence of localised electron states in the form of Ti(III) $3d^1$ ions.[88] Interestingly, in $TiO_2$, polaron formation depends on the polymorph,[83] in particular rutile $TiO_2$ is known to easily form small electron polarons while increasing evidence suggests that anatase $TiO_2$ sustains large electron polarons[89] characterised by shallow states below the conduction band.[90] Such difference in polaron formation can rationalise the distinct behaviour displayed by anatase and rutile in optoelectronic applications.[91] Similarly, polarons have also been reported to impact other oxides used as photocatalysts, electrocatalysts or co-catalyst including $WO_3$,[75,92] $BiVO_4$,[84,93–97] iron oxides,[14,98–101] $Co_3O_4$,[102] and titanates.[103]

**Photoinduced polarons.** Light-initiated charge injection into a semiconductor can also trigger the formation of polarons.[83,104,105] Such photo-induced polarons can be formed indirectly via the injection of electrons from a surface sensitized dye molecule, or directly, though band gap excitation and generation of electron and holes. Mechanistically, photoinduced small polaron formation has been explained with a two-temperature model.[14,106] First, photoexcitation generates a non-thermal distribution of electrons. Such hot distribution thermalizes by creating a population of optical phonons. Finally, the electrons and the optical phonons interact to form the small polaron state (**Figure 2d**).

Photoinduced polaron formation is difficult to track spectroscopically due to the time and length scales involved. It has been studied using X-ray-based techniques taking advantage of their element sensitivity and the ability to monitor changes in oxidation state upon carrier localisation. X-ray absorption spectroscopy monitoring the Fe-K edge, was originally used to detect electron polarons following injection from a molecular dye into $Fe_2O_3$, albeit with limited time resolution.[107] More recently, polarons produced directly via band gap excitation of hematite have been observed using table-top transient extreme-ultraviolet (XUV) spectroscopy on the ultrafast timescale (**Figure 2e**).[14,15,99,108–110]. Such studies have revealed that polaron generation depends on the excitation wavelength, and thus could be related to the wavelength-dependence of photocatalytic activity. Moreover, polaron formation in the bulk of an oxide can occur significantly faster than the equivalent process at the surface, indicating that the coordination environment is critical, and that surface polarons can be controlled by means of materials functionalisation.[106,111] X-ray free electron laser (XFEL) radiation has also been used to track charge localisation processes in $TiO_2$[112,113]; such bright radiation enables new types of experiments, like femtosecond resonant inelastic X-ray scattering (RIXS)[114], that can probe charge carriers and structural dynamics in real time and provide new insights into carrier localisation.

In addition to X-rays, time-resolved microwave conductivity (TRMC)[115] and terahertz conductivity (TRTC) measurements[96] have been used to identify signatures of polarons in $BiVO_4$. Recently, ultrafast spectroscopy with photocurrent detection was employed to detect small electron polarons in $Fe_2O_3$ photoelectrochemical cells during operation.[98] In such device-based experiment, a 400 nm laser pump pulse was used to photo excite an $Fe_2O_3$ photo electrochemical (PEC) cell. Subsequently, a second intense IR beam directly modulated the polaron sub-band gap state and transferred the population of localised states into to higher-lying energy levels. Such re-excitation, drove the system though an alternative pathway which increased the cell photocurrent (**Figure 2b** and **2e**).

While the consequences of polaron formation in photocatalysis are discussed in the next sections, recent methodological advances have sparked a renewed interest in polaron chemistry and we refer the readers to detailed reviews discussing polaronic effects in inorganic and hybrid materials.[83,87,116,117]

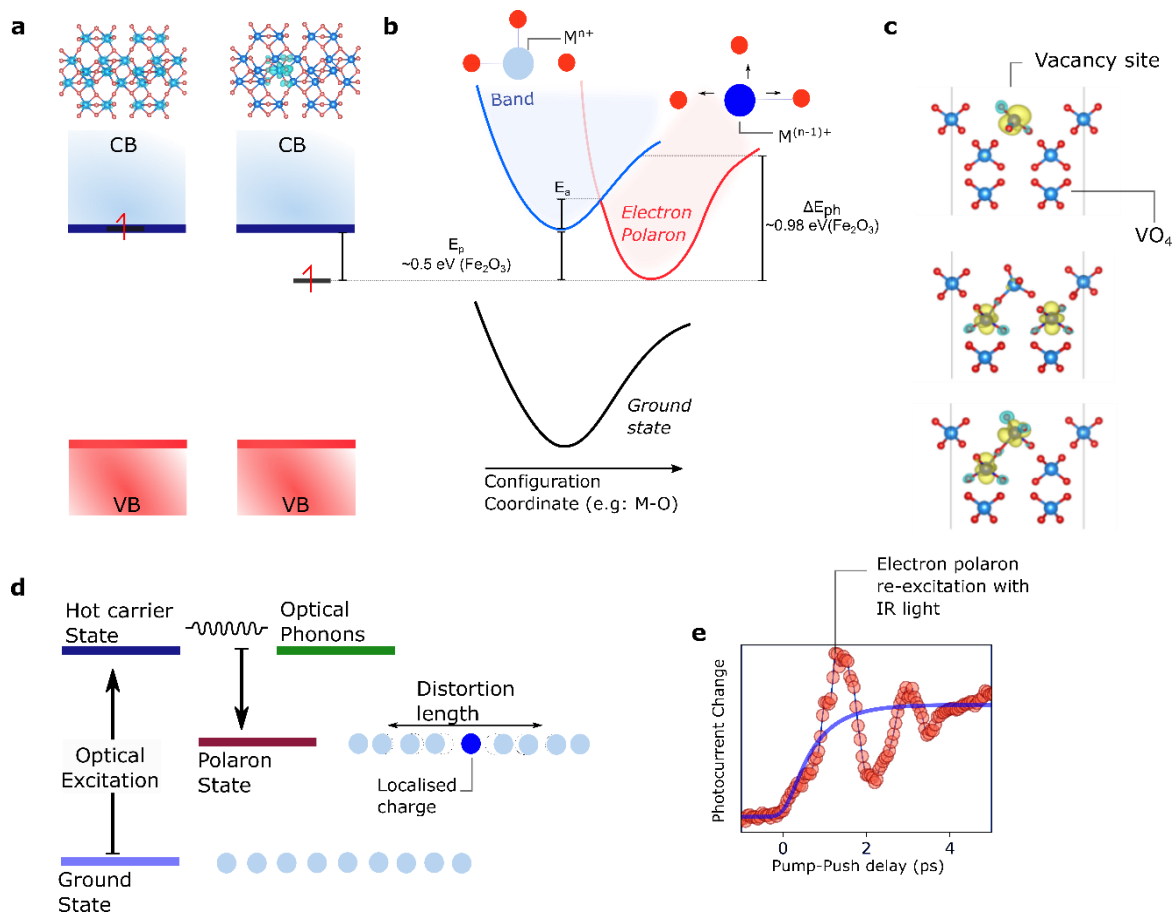

**Figure 2: Polaron chemistry. (a)** A delocalised conduction band electron in $Fe_2O_3$ can localise on an Fe centre, forming a polaronic state which results in an energy level in the band gap. **(b)** Configuration coordinate diagram for polaron formation displaying the polaron stabilisation or binding energy ($E_P$), the activation barrier for polaron formation ($E_a$) and the photon energy required to excite the localised state back into the band state with the same atomic configuration ($\Delta E_{ph}$). **(c)** Different configurations for small polaron formation near an oxygen vacancy on $BiVO_4$. Shown only the V (blue) and O atoms (red). Top: two electrons located at the defect site. Middle: an electron in a corner sharing $VO_4$ and a separate $VO_4$ near the surface. Bottom: electron localised on surface $VO_3$ and corner-sharing $VO_4$. (adapted from:[85]). **(d)** Steps of photoinduced polaron formation upon bandgap excitation. **(e)** Demonstrations of ultrafast photoinduced polaron formation. Pump-push photocurrent data in which the transient polaronic state is re-excited with IR light causing an increase in device photocurrent. The oscillations are associated with coherent phonon generation (adapted from:[98]).

# Photocatalysis with oxides

## Photocatalytic steps

Defects play a key role at each stage of photocatalytic reactions. These can be broken down into five processes as schematically shown in **Figure 3a-e**: (1) Equilibration of the solid and liquid phases and build-up of a semiconductor-liquid junction. (2) Absorption of photons of energy greater than the band gap and generation of charge carriers (electron and holes) that cool down to the band extrema. (3) Charge separation and transport from the point of generation to the surface. Typical minority carrier diffusion lengths ($L_D$) in oxides are in the order of 5-100 nm in oxides.[118] Such distance depends on the carrier mobility (μ) and lifetime (τ) as $L_D = \sqrt{\frac{KT}{e}\tau\mu}$ and is therefore very sensitive to (4) recombination and localisation via defects or polaronic states. (5) Finally, charges must be extracted from the semiconductor.

Except for equilibration with the liquid, all steps are common with other solar energy conversion technologies such as solar cells. However, the charge extraction step imposes severe and unique requirements on catalytic materials. From a band structure viewpoint, extraction in a solar cell involves collection of charges at contact electrodes, whereas the equivalent process in an oxide-based water splitting device is the four-electron oxidation of water. For such selective chemistry to occur, the frontier orbitals must have the right orbital composition and suitable energetics (i.e. in the electrochemical scale E (VBM) > E°(ox/red) for oxidation to occur). From a kinetic viewpoint, while charge collection in a solar cell takes places on the *nanosecond-millisecond* timescale,[119] water oxidation has been measured to occur on the timescale of *seconds* meaning that the system requires $10^3$-$10^9$ times longer-lived carriers as well as mechanisms for their accumulation near the reaction site.[120]

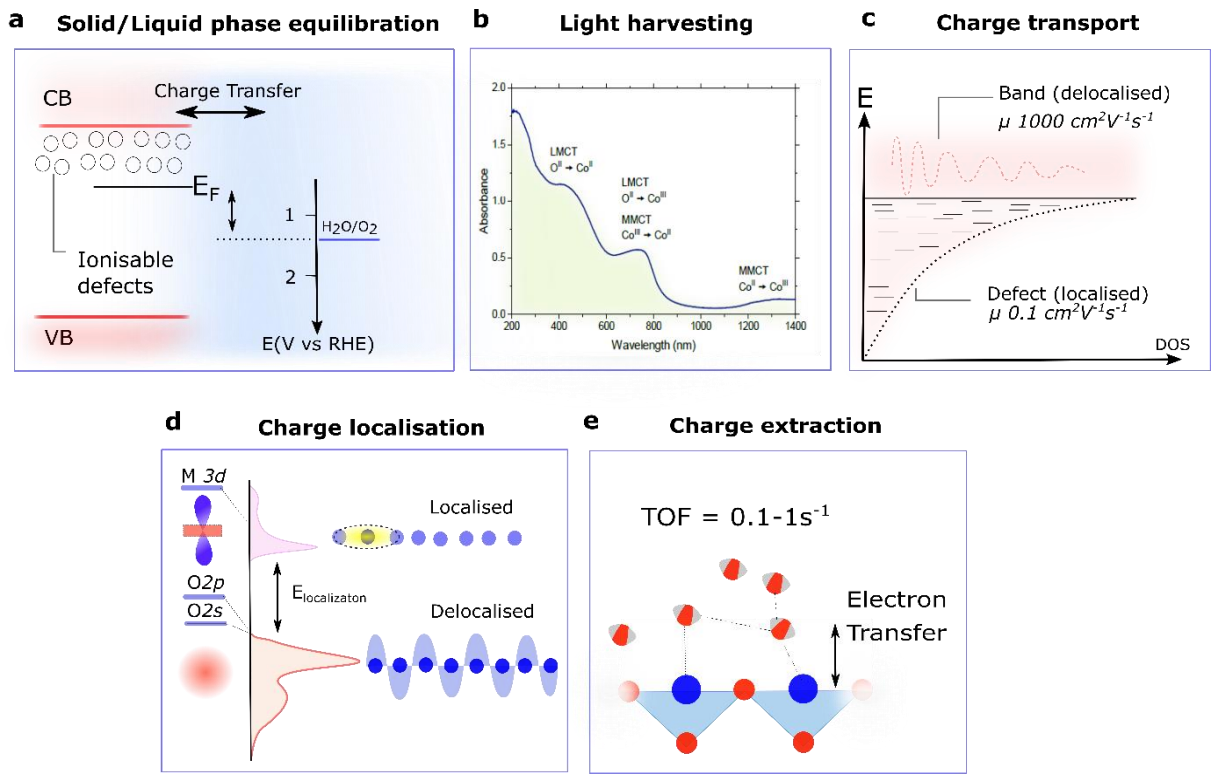

**Figure 3 Fundamental steps in photocatalysis. (a)** Equilibration between the solid and liquid phases. Shown: band positions and thermodynamic reaction energies in RHE scale. In the semiconductor, ionisable donors act as the primary source of charge. **(b)** Absorption of photons for the generation of electrons and holes. Control of the band structure and optical transitions is necessary to maximise absorption in the visible range and ensure the generation reactive charges. The figure shows different type of transitions including Ligand to Metal Charge Transfer (LMCT) and Metal to Metal Charge Transfer (MMCT) in $Co_3O_4$. **(c)** Transport of photogenerated charges. Delocalised band states sustain higher the carrier mobilities (µ) than localised states. (DOS is the density of states) **(d)** Charge localisation in the form of polarons and at physical defects. The nature of the frontier orbitals will dictate the tendency to localise charges. **(e)** Charge extraction from the semiconductor via a surface chemical transformation. Defects and polarons at the surface alter the energy of reactive intermediates and dictate the accumulation of charge carriers at the active site. Turnover frequencies (TOF) typically range from 0.1-1s.

## Impact of defects on photocatalysis

**Electric field build-up.** Semiconductor photocatalysis is enabled by the semiconductor-liquid junction.[121] When the semiconductor is immersed in the electrolyte, charge transfer across the interface takes place until the electrochemical potentials in the solid and the solution are equal, that is until the Fermi level of the semiconductor ($E_F$) and the redox potential of the electrolyte ($E^O$) equilibrate. In the case of an n-type semiconductor with $E_F<E^O$, electrons are transferred to the liquid (**Figure 3a**). The larger number of energy states available in the liquid phase means the energy levels shift mostly within the semiconductor, causing the bands to bend. Through the equilibration process charged defects play an important role by acting as the source of charge. Beyond the surface atomic layer itself, charges depleted away from the surface towards the bulk form a depletion region or depletion layer of width $W_{SCL}$ (**Figure 4a**). The most important consequence of the equilibration process is the formation of a junction that can help charge separation and prevent electron-hole recombination events following illumination. Control over the strength of the interfacial field is possible via changes in the density of ionised dopants ($N_d$), thus via defect engineering, as well as by applying an external potential. The width of the space charge layer is:

$$W_{SCL} = \left(\frac{2\varepsilon\varepsilon_O}{eN_d}\Delta\vartheta\right)^{1/2}$$

(*eq 8*)

where $\varepsilon\varepsilon_O$ are are the material and vacuum permittivity respectively, *e* is the electrical charge, $\Delta\vartheta$ is the difference between a given applied potential and the potential at which the bands are flat and there is no excess charge in the semiconductor (the flatband potential). **Figure 4a**, shows the correlation between the space charge layer width and the density of V(IV)/V(V) states associated with oxygen vacancies in $BiVO_4$.

**Light absorption.** The introduction of impurities has been a major approach to lower the band gap of photocatalytic oxides.[4,122] Doping with metals has been used since the early work on photocatalysis with Cr(III),[123] and V(IV)[124] doped $TiO_2$, as well as in studies of $SrTiO_3$ doped with Mn, Ru and Rh.[4,125] Similarly, the replacement of oxygen atoms with non-metals has been extensively explored and N-doped $TiO_2$ and N-doped tantalates are some of the prime examples where absorption has been enhanced up to 500-600 nm.[126–128] Another common strategy relies on the formation of oxygen vacancies, which has been used to increase the optical absorption of wide gap materials including $ZnO$[129,130], $SnO_2$,[131] $WO_3$[75] and even to produce black $TiO_2$ [132,133] and recently black-$BiVO_4$.[134]

The mechanism by which low dopant concentrations enhance optical absorption is generally the introduction of sub-band gap states that enable longer wavelength optical transitions between defect and band states or between the defects themselves. This is shown in **Figure 4b** for the case of Rh(IV) doped $Sr_3TiO_4$, where the Rh substitutes the Ti.[135] The introduction of $Rh^{4+}$ with configuration $4d^5$ opens up new d-d and charge transfer transitions that enhance the absorption beyond 400 nm.[136] Further doping with La reduces the valence of $Rh^{4+}$ via charge compensation, lowering the absorption in the visible region. This sensitivity to the presence of dopants emphasises that fine control of the intragap density of states is required to modulate the absorption properties on demand. Engineering of the absorption characteristics can be achieved with the principle outlined above, namely: (i) as a first approximation the defect state will have an energy similar to the atomic orbital of origin (e.g. AO of the cation in the case of an anion vacancy) and (ii) the level can be tuned by controlling the coordination and bonding of the defect state.

The power of defect control is seen in the case of $WO_3$ which changes coloration from yellow-transparent to dark blue with increasing density of oxygen vacancies, a property that is exploited for photocatalytic[137,138] and electrochromic applications.[139,140] As shown in **Figure 1e**, when oxygen is released from the material it leaves behind two electrons which localise on two W(VI) centres, reducing them to W(V).[141,142] The resulting intragap energy landscape can potentially be composed of: (i) deep states associated with the actual vacancy (uncoordinated atomic sites); (ii) W $5d^1$ states associated with an electron localised in the vicinity of the vacancy, and (iii) W $5d^1$ states due to an electron and lattice distortion located further away from the vacancy (electron polaron unbound from the source defect).[143] The blue coloration has primarily been attributed to light absorption of W $5d^1$ centres trapped next to the vacancy (bound polarons) and absorption of unbound polaronic states. Both these processes involve charge transfer between W ions, for example from W(V) to W(VI). The spread of defect levels, and thus the optical transitions available, can be tuned by favouring defect-defect interactions at highly reducing growth conditions. It is found that while the defect levels for single oxygen vacancies are located up to 0.7 eV below the conduction band, the formation of defect clusters alters the coordination of the vacancy and its bonding environment. This clustering widens the range of possible charge states and substantially broadens the intragap energy distribution reaching as far as 0.9 eV below the conduction band with a concomitant increase of light-absorption at long wavelengths[75]. Similar defect-defect interactions are known to play an important role in solar cell materials, such as Cu(In,Ga)(S,Se)$_2$,[144] where, in addition, they have been used to control the tolerance to undesired defects. Modulation of defect-defect interactions via concentration and growth environment control offers a strategy to tune density and distribution of intragap states in catalytic oxides.

While enhancing optical absorption is necessary, one critical point to realise is that not all photogenerated charges are able to ultimately drive a desirable photocatalytic reaction. Firstly, charges generated through the direct photoexcitation of defect states, such as oxygen vacancy absorption bands in highly oxygen deficient metal oxides, might lack the necessary driving force to perform thermodynamically demanding reactions.[75] This is shown in **Figure 4c** for WO$_3$ where the absorption enhancement around 750 nm due to defect states does not contribute to the water oxidation photocurrent. Secondly, even charges produced by intrinsic optical transition might not uniformly contribute to catalysis. As illustrated in **Figure 4d,** the average proportion of charges able to drive water oxidation in Fe$_2$O$_3$ PEC cells is substantially smaller than unity. Such phenomena have been ascribed to the formation of photoinduced polarons, which as described below, limit photoconversion efficiencies.[145] The generation of inactive charges is also disproportionately large near the absorption onset, which similarly to observations in early studies, has been attributed to a reduced activity of the ligand field transitions that dominate in this spectral range.[146–148] These observations indicate the need to carefully engineer oxides to ensure that light absorption generates excited states (whether band-like or a polaronic) that can thermodynamically drive catalysis. Moreover, the strong wavelength dependence of photocatalysis highlights the need to better understand how absorption couples to surface catalysis, as well as the intermediate steps such as charge transport and localisation.

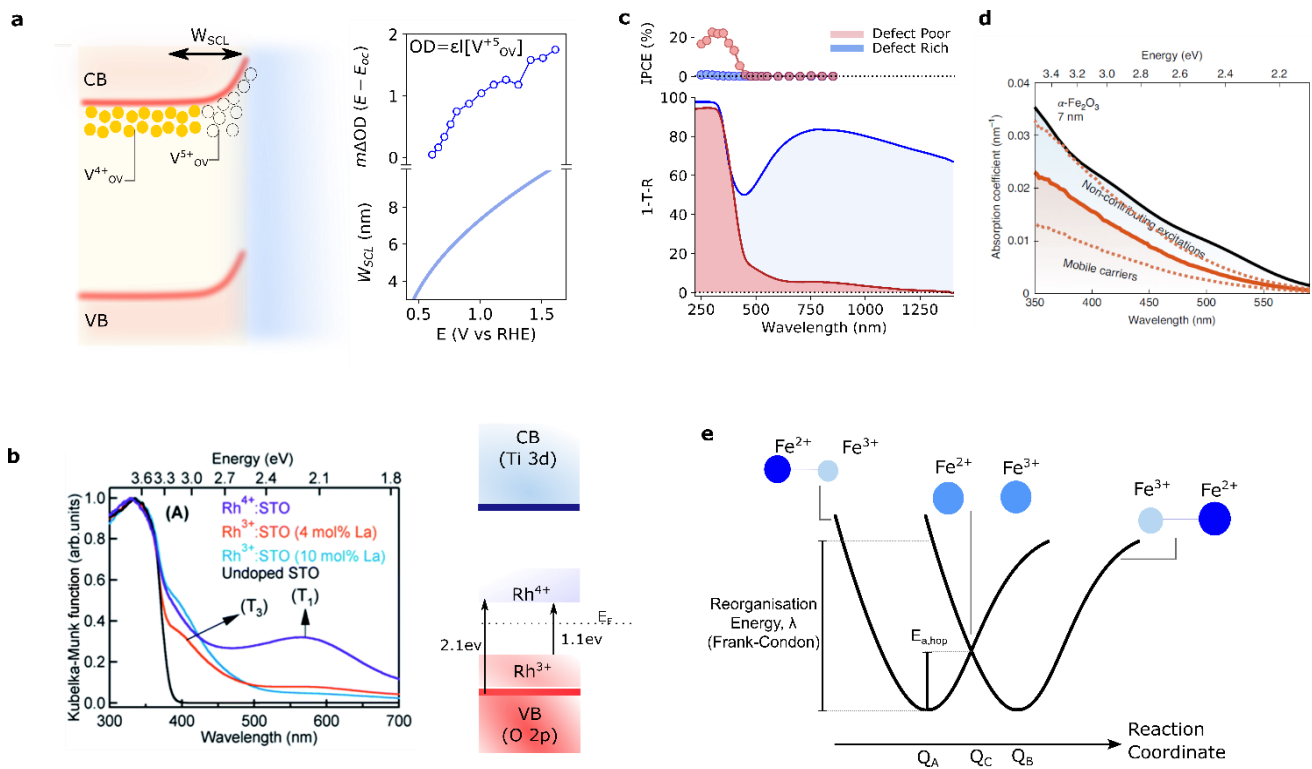

**Figure 4: Role of defects in the solid/liquid equilibration, light absorption and charge transport. (a)** Semiconductor-liquid junction in n-type BiVO$_4$. During equilibration with the electrolyte, filled oxygen vacancy associated states (orange circles) corresponding to V(IV) are emptied (oxidised, empty circles) leading to the formation of a space charge layer of width ($W_{SCL}$). Positive applied potentials, increase the concentration of V(V) states and enlarge the space charge layer (adapted from:[149]). **(b)** Ground state optical absorption of SrTiO$_3$ (STO) measured in diffuse reflectance mode at different doping densities (reproduced from:[136]) and associated band diagram. The diagram exemplifies the role of defect engineering in controlling optical properties of solids. The back trace shows the undoped sample. Doping with Rh$^{4+}$ increases the absorption at long wavelengths (purple). The controlled addition of reduced valence of Rh$^{4+}$ modulates the optical density (orange and blue traces). **(c)** Absorption spectra (1-R-T where R is Reflectance and T is transmittance) of defect poor (red) and defect rich (blue) WO$_3$. The oxygen vacancies increase the optical density in the visible-near IR, however, light absorption at those wavelengths does not contribute to the water oxidation photocurrent as evidenced by negligible IPCE (incident-photon-to-current conversion efficiency) in this range. (Adapted from:[75]). **(d)** Absorption coefficient of a Fe$_2$O$_3$ thin film (black) compared to the estimated upper/lower (dashed orange) and average (full orange) proportion of carriers which are active for water oxidation (adapted from:[145]). **(e)** Configuration coordinate diagram to exemplify the principle of electron-transfer in the small polaron regime. The reactants Fe(II/III) and products Fe(III/II) have equilibrium configurations of $Q_A$ and $Q_B$, respectively. Electron transfer requires a thermally induced event that excites the sample to an activation energy ($E_a$) and brings the system to an intermediate configuration at the crossing point ($Q_C$).

**Charge transport.** One of the main goals for defect engineering of oxide photocatalysis has been to increase the conductivity of majority carriers. For an n-type semiconductor the electron conductivity ($\sigma$) depends on the concentration of electrons (*n*) and their intrinsic mobilities ($\mu$) as:

$$\sigma = en\mu$$

(*eq 9*)

Thus, the introduction of high-valence dopants that create shallow defects increases the conductivity by increasing the electron concentrations in the conduction band. This has been extensively explored, for example, $Fe_2O_3$ has been n-doped with Ti,[150–152] Si,[153] Sn,[154] Ge,[155] amongst others, or $BiVO_4$ has been doped with W and Mo with important implications in their photoelectrochemical performance.[156–158] In addition to metals, the most common source of doping are oxygen vacancies, which are often responsible for the much higher intrinsic doping densities of metal oxides compared to elemental semiconductors.[69]

The underlying doping mechanism is not always fully understood, in particular when the excess carriers form polarons. In the small polaron regime, change transport occurs via hopping between adjacent atomic sites as opposed to band-like transport.[83] This leads to significantly lower carrier mobilities (e.g. ~0.1 $cm^2$ $V^{-1}$ $s^{-1}$ of $BiVO_4$) than in traditional semiconductors (e.g. ~1350 $cm^2$ $V^{-1}$ $s^{-1}$ for electrons in Si).[159,160]. The reason for such slow transport is that movement of a polaronic charge from site to site requires a structural rearrangement with an activation barrier with $\mu_{pol} \alpha\ e^{-E_{a,hop}/KT}$ .[161,162] In order for the electron transfer to occur, the reactants must first be excited (e.g. thermally) to an activation barrier and achieve an intermediate configuration, as schematically shown in **Figure 4e**. Note that in the diabatic regime, the activation barrier for hopping is related to the polaron binding energy as $E_P \sim 2E_{a,hop}$[98,161] and thus, the more strongly bound the polaron, the lower its mobility.

$BiVO_4$ is an example of a photocatalyst where electron transport is considered a limiting factor.[149,163,164] While oxygen vacancies are considered the be the source of n-doping, several spectroscopic studies have found that vacancies give rise to donor states as far as 0.6 eV below the conduction band.[149,164] Such deep character should prevent these states from being ionised and thus should render them ineffective n-dopants. However, recent theoretical work has explained this discrepancy suggesting that while oxygen vacancy states are indeed deep, their ionisation to more conductive polaronic states is only in the order 0.1-0.3 eV.[165] This vacancy-to-polaron ionisation value is consistent with experimentally measured activation energies for carrier de-trapping.[149] Polaron transport is also impacted by the differences in the concentration of oxygen vacancy between bulk and surface, as observed previously using transmission electron microscopy,[166] as well as by variations in the adopted geometrical distortions and configurations[167] all of which can significantly alter the mobility and energetics of the respective polarons.[85]

In addition to the effects of vacancies, the presence of other imperfections in the crystal can also change polaron transport. For example, doping of $Fe_2O_3$ with Si has been reported to increase the conductivity by increasing carrier concentrations but also by enhancing the mobility of small polarons.[100] The incorporation of Si was found to enhance transport by changing bond lengths, altering hoping probabilities in the system and lowering the activation energy for hopping (**Figure 4e**). Critically, such positive effects of Si are not general for all dopants and might depend on the radius of the foreign atom and the extent to which it distorts the crystal. For example, Sn-doping was found to have a negative effect on polaron mobilities, increasing the activation barrier for hoping.[153] However, Sn-doping is actually used in some of the best performing $Fe_2O_3$ photoelectrocatalysts as it is able to compensate its detrimental effect on polaron transport by enhancing electron concentrations. These results evidence that doping is more a more complex process than injecting carriers into a band. The countering effects that atoms such as Si and Sn exhibit might prove useful to identify an optimum of carrier densities and polaron transport via careful co-doping strategies.

Recent pump-push photocurrent experiments of $Fe_2O_3$ PEC cells have also demonstrated that electron polaron formation occurring within less than 1 ps impacts charge extraction in the photoelectrochemical device as there is a large population of polaronic charges that recombine before being extracted.[98] Similarly, ultrafast terahertz conductivity measurements in $BiVO_4$ [96] have reported a combined electron-hole peak mobility of 0.4 $cm^2$ $V^{-1}$ $s^{-1}$ at ~1 ps that decreases by ~70% within the first few hundreds of picoseconds. This loss signifies a substantial loss of free or relatively unbound carriers, which can result from carrier localisation. Time-resolved microwave conductivity measurements on the nanosecond – millisecond timescale have yielded similar low mobilities for un-doped $BiVO_4$ further supporting the localised picture.[115] However, the observed carrier lifetimes of ~40 ns lead to relatively long calculated carrier diffusion length of ~70 nm, and suggests that despite the low mobility, photogenerated charge carriers are able to access interfaces for reactions in this material.

**Charge localisation on defects.** Even when a beneficial dopant is identified, low doping efficiency is a problem that can affect many oxides. This happens when only a small fraction of the dopants are active (ionised). It can occur either from dopant levels being too deep in the band gap or due to the migration of dopants to crystal boundaries where their behaviour changes and they become inactive.

Deep defect states act as trapping sites for photogenerated charge carriers. Such trapping process has two main consequences: (1) to immobilise one type of charge carrier via selective trapping and (2) to enable non-radiative electron-hole recombination via a Shockley–Read–Hall mechanism. The first process can actually be beneficial as trapped carriers can exhibit very long lifetimes, as observed in different polymorphs of $TiO_2$ as well as in $Fe_2O_3$ or $WO_3$[120,168–171]. In the absence of other pathways to increase carrier lifetimes, they can aid charge separation and ultimately improve catalysis.

The complexity of the localisation process is exemplified in **Figure 5a** for $WO_3$. In $WO_3$ a large part of the W 5*d* vacancy state distribution is energetically too far from the conduction band edge for thermal excitation.[75] As a consequence, intragap states in $WO_3$ are largely occupied with electrons (in the form of reduced W(V) centres) and thus can act as trapping sites for minority carriers (photogenerated holes). Such trapping enables a relaxation path for the hole in which it lowers its energy by oxidising a W(V) centre, losing most of its oxidative driving force in the process. Critically, the hole trapping does not immediately lead to annihilation of the counterpart photogenerated electron. This is corroborated by measurements on longer timescales showing a substantial gain in lifetime of up to several milliseconds[75]. Hence the localisation process provides a mechanism to extend carrier lifetimes at the expense of carrier energy or driving force.

Interestingly, the process can be tuned by controlling the distribution of intragap defect states and, as expected, highly oxygen-deficient $WO_3$ with a broad defect distribution exhibits faster trapping than the near-stoichiometric analogue. This is exemplified in the transient dynamics of $WO_3$ shown in **Figure 5b** where the appearance of the negative signal is associated with hole trapping.[75,171] The same behaviour has been observed in $BiVO_4$ where within less than 200 fs of photoexcitation, a fraction of the photogenerated holes relax into intragap V(IV) states associated with oxygen vacancies, transforming them into the V(V). This relaxation into such states located ~0.6 – 0.8 eV relative to the conduction band and thus more negative than the thermodynamic water oxidation potential of 1.23 $V_{RHE}$, renders these photogenerated holes inactive for water oxidation.[149] It is important to note that while a decrease in driving force associated with carrier relaxation means that the charges might become unreactive towards highly demanding, multi-electron reactions, the concomitant increase in lifetime might be advantageous to drive less energetically demanding processes such as pollutant degradation. Hence, engineering of defect distributions and defect concentrations might open up a route to tune oxides to specific catalytic reactions. In this sense, thermodynamically facile reactions might be more tolerant and even benefit from high defect concentrations while more demanding processes might require lower defect densities. Indeed the performance optimum for water oxidation in $WO_3$ was found at only ~2% of oxygen vacancies.[12]

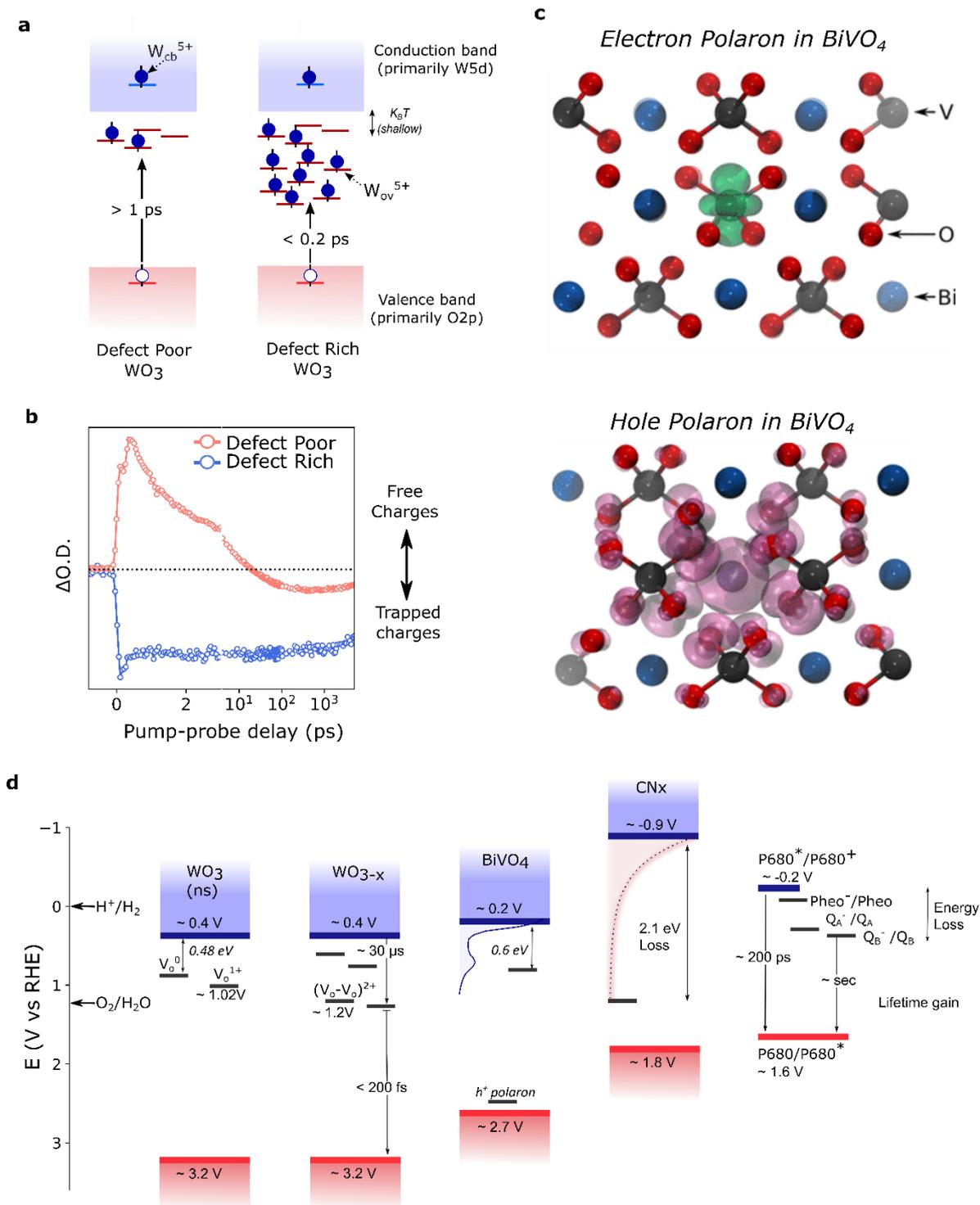

**Figure 5: Charge localisation on defects and polarons. (a)** Band diagram of WO$_3$. Deep donor states are occupied with electrons, providing a pathway for hole trapping. **(b)** Kinetics of ultrafast hole trapping or localisation into W$^{5+}_{ov}$ associated with oxygen vacancies in WO$_3$, where negative signals are associated with trapped holes. The localisation process is near-instantaneous at high vacancy densities (adapted from:[75] ). **(c)** Isodensities of the electron and hole polarons in BiVO$_4$. The electron localises strongly on the V atom changing the oxidation state from V(V) to V(IV) and increasing the V-O length. The hole polaron distributes between the Bi

and O atoms slightly shortening the Bi-O bond (adapted from:[84]). **(d)** Intragap defect levels in catalytic oxides, carbon nitride and redox-cofactors in photosystem II shown relative to the water splitting thermodynamic potentials.[75,84,172–174] The shaded area in $BiVO_4$ corresponds to the distribution of $V^{4+}/V^{5+}$ obtained from capacitance measurements.[149] In PSII, charge separation takes places via a series of redox cofactors which enable the necessary lifetime gain for water oxidation to take place at the expense of a major energy loss. Similarly, in artificial systems photogenerated hole trapping into deep defect states, results in an increase of carrier lifetimes at the expense of a loss in oxidative power. In $WO_3$, '*ns*' indicates near stoichiometric.

**Charge localisation on polarons.** The formation of polarons upon photogeneration provides another pathway for energy minimisation. In $BiVO_4$, hole polarons are reported to be weakly localised and locate only 0.1 eV above the VBM, consequently polaronic holes can drive water oxidation. In contrast, electron polarons are strongly localised on V atoms and exhibit stabilisation energy as large as 0.9 eV relative to the CBM (**Figure 5c**).[84] The electron polaron level is more positive than the $H^+/H_2$ redox potential rendering polaronic electrons unreactive towards proton reduction and limiting the ability of $BiVO_4$ to drive the overall water splitting reaction. Moreover, the electron level is positioned close to the $H_2O/O_2$ level, indicating that recombination of electron-hole polarons can effectively compete with the charge transfer to the electrolyte.[84] Note that polaron binding energies depend on temperature, the presence of other defects and the physical location within the oxide (surface/bulk). Consequently, there is spread of reported absolute values reported in the literature.

In $Fe_2O_3$, the stabilisation energy for electron polarons is also large (~0.5 eV) and reduces the optical gap (2.2 eV) to an effective electronic gap of only~ 1.75 eV.[98] Recent X-ray photoemission studies have suggested that such strong reduction in the energy of the electronic state limits the attainable photovoltage in hematite devices and thus restricts the ability to drive photochemical reactions.[175] Similar to what has been observed in other systems such as organic semiconductors, it might be possible to reduce such losses via polaron engineering. One route could be to reduce the binding energy or increase the activation barrier for polaron formation through crystal engineering (**Figure 2b**). This would decrease the probability of polaron formation and minimize the reduction in the energy of the electronic state easing the penalty on the photovoltage. Moreover, lower polaron binding energies would also induce lower activation barrier for carrier hopping, increasing transport. Guiding the re-design of polaronic extended solids is non-trivial, but classical concepts derived from coordination chemistry might be of use. Indeed recent experiments have shown that photoinduced polaron formation depends on molecular-scale parameters such ligand field strength and metal coordination environment.[108] In this context, strategies like the use of metal-oxyhydroxides or multinary oxide systems such as ferrites, currently being explored for photocatalysis[176] could help control polaron impact.

The band alignment of several commonly used oxides and their respective intragap states are shown in **Figure 5d**. While it is established that defects impact the performance of these oxides, it is yet an open question whether charge carrier localisation process can be overcome to increase catalytic yields or if they are unavoidable, and potentially even necessary, to achieve sufficient lifetime-gains. Interestingly, the trade-off between lifetime gain and energy loss observed in oxides is not unique to solids but is a central feature in the operation of the natural photosynthetic system. In Photosystem II, a series of redox co-factors enable the spatial separation of the chlorophyll excited state.[172] The result is a lifetime gain of $10^6$ at the expense of an energy loss of 700 meV, equivalent to 40% of the energy of a red photon.[177] Such co-factors states can be approximated as the 'molecular analogue' of intragap states in extended solids. However, the photosystem can modulate the redox cofactors to regulate the backward and forward electron transfer and thus displays significant adaptability and photoconversion flexibility.[178] In order to achieve high photocatalytic efficiencies, our challenge is to outperform nature by reducing or better controlling the energy losses in artificial systems. Learning to minimise unwanted defects and how to energetically and spatially control defect populations could enable the necessary fine-tuning and even expose new materials functionality.[179,180] Achieving this goal might require exploiting defect-defect and polaron-defect interactions to tune energy distributions and the tolerance to unwanted defects.

**Surface chemistry**. The last stage in the photocatalytic process is the extraction of charges from the active material. This final step marks the start of the catalytic part of the mechanism in which photogenerated charges are extracted via a chemical reaction occurring at the solid/liquid interface. Many reactions of interest such as proton reduction or water oxidation are multi-electron reactions and involve the formation and accumulation of several reactive intermediates, which means that the completion of the catalytic cycle and final product formation are slow (milliseconds to seconds).[181,182] The role that defects play at this stage is challenging to establish. Experimentally, it is difficult to characterize the surface under operation to elucidate, at an atomic level, how the surface changes upon contact with the electrolyte or under different working conditions. This is not a minor issue, as for example, in oxide photoanodes the OER mechanism has been proposed to change as a function of pivotal parameters like the illumination intensity.[183,184] From a simulation viewpoint it is difficult to build realistic models that describe the influence of solvation and temperature fluctuations on the frontier orbitals due to the large computational cost. Such information is essential to explain reactivity and defect formation at the interface. Consequently, the surface of photocatalysts under working condition is, to a great extent, uncharted territory. Nonetheless, new experimental methods and computational electrochemistry approaches are increasingly enabling insights into the solid/liquid interface.[82,185–192]

One of the biggest consequences of the lack of operando surface characterization is the uncertainty in the energy levels of defects, polarons and even band edges in contact with the liquid. This means that in many cases it is even unclear if a given state can thermodynamically participate in redox chemistry. This has led to contrasting interpretations of the role of defects, with some reports suggesting that defects have positive effects and mediate catalytic reactions, while others propose that defects are unreactive, detrimental and should be eliminated or passivated.[193–196] From a reactivity standpoint, the basic roles that defects play in surface chemistry are schematically represented in **Figure 6a-d** and can be categorized as follows: (a) the defect forms as part of the catalytic cycle and interaction with the electrolyte.[197,198] (b) The presence of defects 'physically' changes the nature of the surface by, for example, exposing new reactive sites.[199] (c) The defect modulates the availability of electronic charge at the surface and the charge transport between surface sites[85,92,167] and (d) the presence of defects modulates the energetics of the surface by, for example, altering the energy of reaction intermediates, facilitating adsorption of molecules and even modifying the magnetic properties of the surface.[200–204]

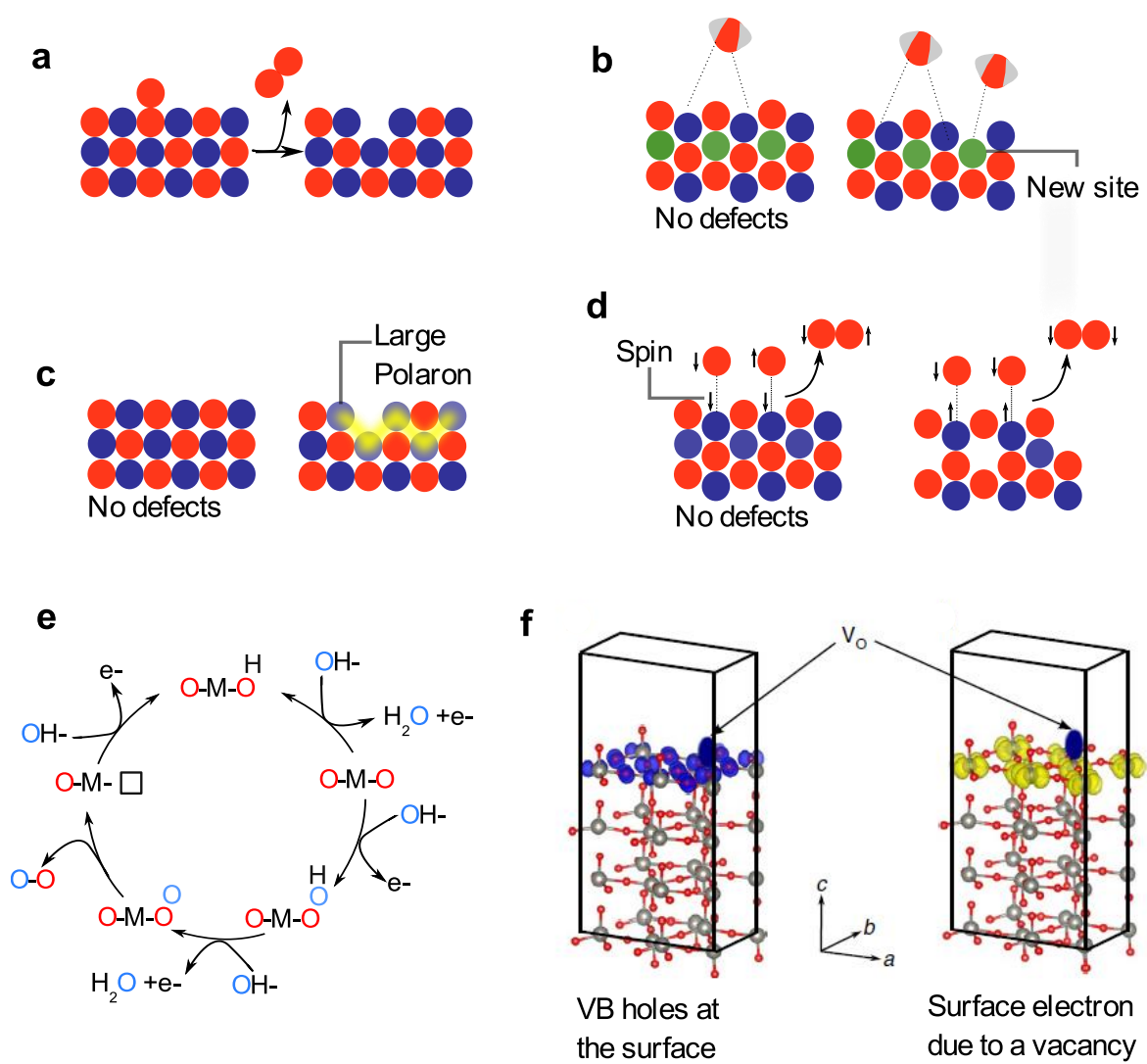

**Figure 6: Role of defects on the surface catalysis. (a)** Formation of an oxygen vacancy during the catalysis. During the oxygen evolution reaction (OER), the evolved $O_2$ might contain a lattice oxygen. **(b)** The formation of defects can expose new active sites which help enhance catalytic activity. **(c)** Excess charge associated to defects can localise at the surface forming small and large polarons. Such localisation will dictate the ability of charge to diffuse and accumulate on one site. **(d)** The formation of defects can alter the surface properties by for example changing the surface spin polarisation and favouring the parallel alignment of spins. This might impact spin-restricted reactions. **(e)** A possible mechanism of the OER involving the formation of an oxygen vacancy. The evolved $O_2$ contains water and lattice oxygen atoms . **(f)** Valence band holes (blue) and excess electrons (yellow) associated with a vacancy (Vo) at the surface of $WO_3$. The holes at the valence band maximum originate from O 2$p$ orbitals while the electrons originate from W 5$d$ orbitals. Atoms: oxygen (red), tungsten (gray). Adapted from:[92].

Given the abundance of the oxygen atom in oxides, the formation of oxygen vacancies is particularly important, especially in water-based electrolytes. In perovskite electrocatalysts, $^{18}O$ isotope labelling mass spectrometry has shown that the $O_2$ evolved during the OER can actually come from the lattice as opposed to exclusively from the water (**Figure 6e** shows a potential mechanism).[205] The activation of lattice oxygen and vacancy formation during the catalytic cycle (with their subsequent 're-filling' ) has not been extensively studied in photocatalysis, but is a potential route to increase the redox activity. As expected from the discussion above, whether vacancies are generated or not will depend on the metal-oxygen bonding and the nature of the frontier orbitals. Such effects have been systematically studied in oxide electrocatalysts in which it was found that higher bond covalency increases the O 2$p$ character near the fermi level and favours the participation of lattice oxygen atoms during catalysis.[205,206] Control over the covalency and increase of the overlap between metal and oxygen orbitals can be achieved by reducing the energy of the cation (more electronegative metal) or incorporation 4$d$ cations with more spatially extended orbitals (**Figure 1c**).[197] The critical aspect is that in these covalent systems the band structure allows the creation of holes in the oxygen band, thus enabling vacancy formation. This might also be the case in oxide photocatalysts. For example, recent photoelectron spectroscopy studies on $BiVO_4$ have shown that upon exposure to water the V and Bi atoms are reduced and the oxygen content at the surface decreases due to the formation of oxygen vacancies. [198] Such findings emphasize the importance of oxygen vacancies at the solid/liquid interface and underscore the need for surface characterization and labelling studies to expose the role of lattice oxygen in photocatalytic reactions.[207–211]

In addition to their participation in the catalytic cycle, oxygen vacancies in $BiVO_4$ have also been shown to alter the surface reactivity towards the OER. This has been proposed to occur via two mechanisms, namely by activating the V atoms, and thus increasing the number of active sites, and by changing the energy of adsorbates and enhancing hole transfer.[199] Moreover, the presence of vacancies also increases excess charge at the surface with potential implications for reactivity. In $WO_3$, this surface defect has been calculated to be favourably positioned to transfer electrons to OH, possibly forming $OH^-$. Subsequently, delocalized surface holes might transform the hydroxide into a reactive $OH^*$ intermediate. Critically, in $WO_3$, the excess charge due to the vacancy has been proposed to form a delocalized large polaron (1nm in diameter, **Figure 6f**).[92] Control over polaron formation at the surface is critical, as for example, recent measurements have shown that the rate-determining step in the OER on $Fe_2O_3$ involved a triply oxidized cluster.[212] The accumulation of three oxidation equivalents on one specific site requires the lateral diffusion of holes over the surface, and consequently necessitates of an efficient transport mechanism (note that a strong polaron localization with high binding energies also leads to a higher activation barrier for charge hoping as discussed above).

The changes that defects such as dopants or vacancies have on the surface chemistry can be more intricate than adding excess charge or exposing new sites. For example, recent experimental work has shown that tuning the concentration of Ti vacancies in $TiO_2$ provides a path to alter the spin polarization in the system and enhance photocatalytic hydrogen evolution and phenol degradation.[202] It was found that an increase of Ti vacancies increases the number of electrons in spin-down sates. Such electronic polarization favours the formation of electron (holes) with spin-down upon photoexcitation and was suggested to influence the recombination of hydroxy radical intermediates via spin-restriction processes (i.e. a spin down hole can accept electrons from solution $OH^-$ species which have also the same spin direction). Such magnetic effects are not exclusive to $TiO_2$; the use of magnetic fields was recently proposed to enhance the OER Ni-Fe oxide electrocatalysts by favouring the spin parallel alignment of oxygen radicals in the O-O bond formation.[213] These advances suggest that tuning the spin polarization of the active layer, via magnetic fields or defect formation, could provide a new avenue to boost photocatalytic reactions.

## **Learning from other photocatalysts**

Although this review focuses on metal oxides, an awareness of defect states is important for any photocatalyst. We therefore end with a brief comparison to defects in other photocatalytic materials to put the defect chemistry of oxides into perspective.

**Organic photocatalysts.** Organic polymers have recently become popular photocatalysts because of their high synthetic tunability and are emerging as promising photocatalyst candidates. This high synthetic tunability has sparked the creation of a large variety of polymeric materials with varying degrees of light absorption, conjugation, porosity and hydrophilicity amongst other factors.[214–216] The most prominent examples of polymeric photocatalysts are carbon nitrides $CN_xH_y$, which are based on triazine or heptazine units and possess band gaps of 2.7 eV[217]. Like metal oxides, carbon nitrides have been explored for a number of catalytic reactions such as $H_2$ evolution, $O_2$ evolution, $CO_2$ reduction and pollutant degradation.[218]

While the nature of the frontier orbital is clearly different to that of oxides, increasing evidence suggest the activity of newly developed carbon-nitride photocatalysts is also strongly impacted by their defect chemistry in a similar way to oxides. Spectroscopic studies indicate that charge trapping in urea-derived materials results in energy losses of about 1.5 eV (~ half the energy of absorbed photons), significantly reducing the reactivity of photogenerated charges and suggesting that deeply-trapped long-lived electrons are unreactive for photoreduction (**Figure 5d**).[174,219] On the other hand, in melamine derived carbon nitride, long-lived trapped electrons were found to retain sufficient chemical potential to drive proton reduction even after 12 hours in the dark.[220,221]

Another similarity of organic semiconductors with metal oxides is the tendency to form polarons upon photoexcitation. This has been observed in organic photocatalyst and especially in photovoltaics. In general, polarons are known to be detrimental to the dissociation of charge transfer states and consequently reduce the voltage output of devices. This parallels the behaviour proposed for metal oxide PEC cells. What is most interesting is that, experimentally, it is found that efficient organic systems do not actually require band states for charge separation, indicating that the detrimental influence of polaronic effects can be reduced by material engineering.[222] Similar synthetic strategies to those developed for organics could potentially be applied to oxide photocatalysts in order to circumvent polaron-associated losses.

Despite the big difference between carbon-based and inorganic photocatalysis, it is apparent that they share similar molecular-scale principles in relation to defect chemistry. These similarities offer an exciting opportunity to draw comparisons between these systems and, for example, use the synthetic tunability of organics to devise new defect control strategies that are also applicable to oxides.

**Metal halide perovskites.** Organic-inorganic (e.g. $CH_3NH_3PbI_3$) and total inorganic (e.g. $CsPbI_3$) halide perovskites are increasingly being explored for photocatalytic applications.[223–227] In comparison to metal oxides, these systems exhibit higher conduction band ($E_{CBM}$< the $E^o(H_2/H^+)$) and valence bands energies ($E_{VBM}$< $E^o(O_2/H_2O)$) and, consequently, are explored for reduction reactions or thermodynamically facile oxidation of organic compounds.[228–233]

Generally, metal halide perovskites exhibit relatively low defect formation energies and thus sustain large defect concentrations. However, they are heavily compensated systems where the large concentrations of positively and negatively charged defects cancel resulting in low electronic carrier concentrations. They have also proven very difficult to dope, as these soft crystals can self-regulate their charge by easily forming compensating defects. Another difference to oxides is that such defects tend to be shallow in nature and thus relatively benign in the halides.[234,235] This is a consequence of the electronic structure which significantly differs from oxides, in addition to the low carrier effective masses and the large dielectric constants owing to their polarisable structure.

The 'defect-tolerant' electronic structure of halide perovskites means that photoexcited charges retain most of their electronic energy, a positive feature for photocatalysis. While there is still no consensus on the exact mechanism, increasing evidence indicates that (large) polaron formation assists screening of the excited states and helps reduce carrier recombination rates.[236] The formation of large polarons, as opposed to small polarons with large binding energies typical of oxides, is due to the stronger orbital overlap that results in lower effective masses.[117,237–239] Most critically, polaronic effects can be engineered as the softness of the material can be modulated by varying chemical composition.[240] Metal halide perovskites are an example where the trade-off between energy-loss and lifetime-gain, characteristic of photocatalytic oxides and the natural photosynthetic system as discussed above, can be more readily optimised.

The major limiting factor of metal-halide perovskites for photocatalysis is their poor stability, especially in polar solvents and under illumination. For example, when exposed to water the material tends to sufferer surface reconstructions, forming grain boundaries. Mixed halide perovskites tend to undergo phase separation under constant illumination,[241] and the facile formation of iodine vacancies can promote degradation as the excess charge associated with the vacancy reacts with $O_2$ to form superoxide. However, further studies are required to unveil the role of defects in photocatalytic perovskites as this might differ from their role in solar cells. For example, recent work on organic oxidation has shown that iodine vacancies are actually the central species dictating photocatalytic activity.[229]

## **Outlook**

A big part of the challenge facing photocatalysis involves balancing the supply and demand of reactive charges by coordinating the charge generation and separation steps with an efficient charge extraction. Kinetically, this requires preparing charge carriers that live long enough to supply a catalytic active site and support high turnover numbers. Energetically, such long-lived carriers must retain sufficient free energy to drive the desired chemical transformation. In a solar cell this synchronization is relatively simple and in Photosystem II it is achieved through a series of complex and dynamic structures within the protein frameworks. While such structures are difficult to replicate synthetically, defect engineering might provide the chemical complexity and tunability needed to improve artificial systems.

From a materials viewpoint, optimizing the trade-off between lifetime-gain and energy-loss will require: (i) going beyond current doping strategies and learning to control sub-band gap states using defect-defect and defect-polaron interactions and (ii) exploring polaron engineering strategies in oxides by synthetically tuning the metal coordination environment and altering the structural softness. Experimentally, such advances will require (iii) the in-situ characterisation of defects and polarons during operation and addressing how the complex energy landscape that defects generate is affected by the catalytic conditions. Such studies should be guided by (iv) detailed theoretical work that pushes beyond the current standard of studying isolated defects in a perfect host at 0 K and explores systems at room temperature, equilibrated with liquid electrolytes and that account for the multiple configurations that vacancies and other defect species can take, both at the bulk and at crystal boundaries.

**Box 1: Solid-State Photocatalytic approaches**

Solid-state photocatalysis is generally achieved using particulate systems in suspension and films or via photoelectrochemical (PEC) cells (**Figure B1a**-c).[3,242,243] In a particulate system (**Figure B1a**), a single particle drives a complete redox reaction (oxidation and reduction), often with the aid of a co-catalyst. In such systems catalytic particles are in close proximity, limiting solution resistance and mass transport losses but also complicating the collection and separation of products. While traditionally this suspension photocatalysis had been performed in 'baggie systems', recently this technology has been implemented in the form of photocatalytic panels in which the particles are immobilised on support panels. In contrast to suspension systems, in PEC cells each reaction takes place at a separate electrode – oxidation at the (photo)anode and reduction at the (photo)cathode – thereby facilitating product separation and collection (**Figure B1b**). There are multiple PEC cell configurations,[244,245] for example, the electrodes can be joined together forming a compact device or they can be physically separated from one another and connected via wires. Equally, some systems employ two semiconductors as photocathode and photoanode, while other systems employ just one photoelectrode and a second metallic electrode. The latter case requires the semiconductor to generate enough photovoltage to drive the full reaction (oxidation plus reduction) or relies on the assistance of an external voltage source, for example a solar cell. It is important to note that, although the eventual goal is to design a complete functional system, the oxidation and reduction reactions are usually studied separately in order to optimise active materials. This is achieved by using sacrificial donors in suspensions or by voltage assisted PEC cells involving the desired photoelectrode and a metallic counter electrode (**Figure B1c**).

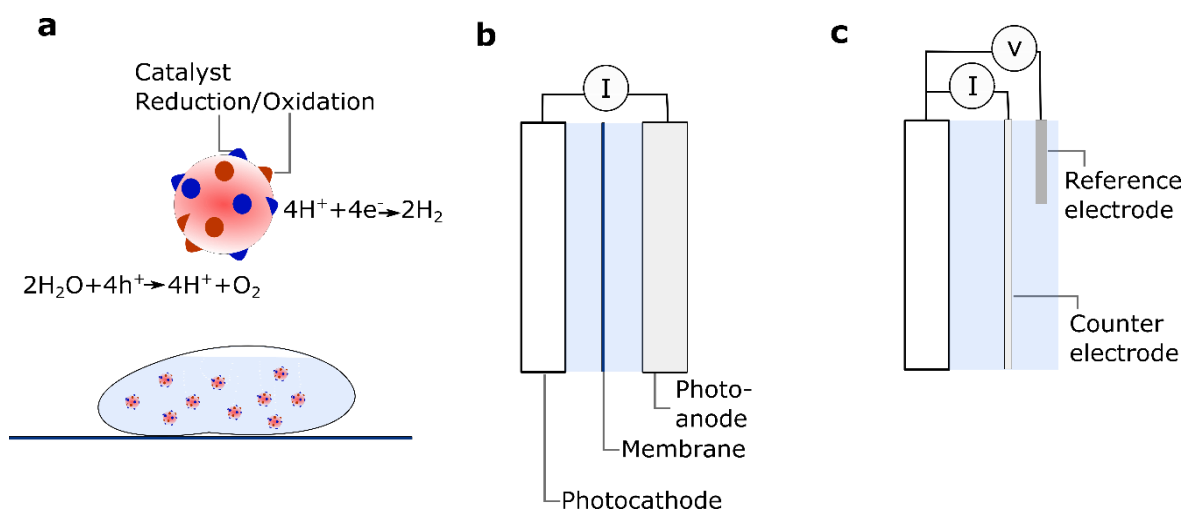

## Acknowledgments

We thank Liam Harnett for sharing his data and insights on $WO_3$. E.P acknowledges the support form IJC2018-037384-I funded by MCIN/AEI /10.13039/501100011033. S.S. and J.D. acknowledge funding from the European Union's Horizon 2020 research and innovation programme under grant agreement 732840-A-LEAF. M.S. thanks EPSRC for a Doctoral Prize Fellowship.